\documentclass[aps,prl,reprint,showpacs,floatfix,nofootinbib]{revtex4-1}

\usepackage{bm}
\usepackage{graphicx}
\usepackage{amsmath}
\usepackage{natbib}
\usepackage{color,soul}

\DeclareMathOperator{\sinc}{sinc}

\begin{document}

\title
{
	Effects of Dephasing on Spin Lifetime in Ballistic Spin-Orbit Materials
}
\author
{
	Aron W. Cummings$^{1}$ and Stephan Roche$^{1,2}$
}
\affiliation
{
	$^1$Catalan Institute of Nanoscience and Nanotechnology (ICN2),
	CSIC and The Barcelona Institute of Science and Technology,
	Campus UAB, Bellaterra, 08193 Barcelona, Spain
	\\
	$^2$ICREA, Instituci\'{o} Catalana de Recerca i Estudis Avan\c{c}ats,
	08070 Barcelona, Spain
}
\date{\today}

\begin{abstract}

We theoretically investigate spin dynamics in spin-orbit-coupled materials. In the ballistic limit, the spin lifetime is dictated by dephasing that arises from energy broadening plus a non-uniform spin precession. For the case of clean graphene, we find a strong anisotropy with spin lifetimes that can be short even for modest energy scales, on the order of a few ns. These results offer deeper insight into the nature of spin dynamics in graphene, and are also applicable to the investigation of other systems where spin-orbit coupling plays an important role.
\end{abstract}


\maketitle

\textit{Introduction}. Following the description of Rashba spin-orbit coupling (SOC) in two-dimensional electron gases (2DEGs) \cite{RashbaSOC}, understanding the spin dynamics in these systems has been essential for proposing spintronic devices \cite{DattaDas} and predicting fundamental physical phenomena \cite{SHE1, SHE2, SHE3}. Rashba SOC allows for the electrostatic manipulation of spin states, paving the way towards non-charge-based computing and information processing \cite{RashbaReview1}. Beyond traditional semiconductor quantum wells, 2D materials including graphene and ${\rm MoS_2}$ monolayers have generated significant interest. In addition to their predicted long spin lifetimes \cite{theory1, theory2, theory3, mos2_1, mos2_2}, the possibility to harness proximity effects or to couple the spin and valley degrees of freedom makes these materials interesting both fundamentally and technologically \cite{mos2_3, mos2_4, graphene_prox1, flagship}.

From a practical perspective, understanding spin lifetimes in clean materials is a prerequisite to realizing spintronic devices, since they determine the upper time and length scales of operation. In Rashba SOC materials, the spin lifetime is normally dictated by the Dyakonov-Perel (DP) mechanism \cite{dp1}, where SOC induces spin precession of charge carriers. After many scattering events the randomization of precession leads to dephasing and a loss of the spin signal, such that the spin lifetime $\tau_s$ scales inversely with the momentum scattering time $\tau_p$. This contrasts with the Elliot-Yafet (EY) mechanism \cite{ey1,ey2}, for which charge carriers can flip their spin upon scattering, giving $\tau_s \propto \tau_p$. The EY mechanism usually dominates in disordered metals, but its contribution has been also discussed for graphene \cite{ey_graphene}.

The SOC in graphene is predicted to be small, on the order of $\mu$eV \cite{soc1,soc2,soc3,soc4,soc5}, leading to estimates of $\tau_s$ in the micro- to millisecond range \cite{theory1,theory2,theory3}. In contrast, experimental spin lifetimes range from hundreds of ps to a few ns for non-local Hanle measurements \cite{hanle1,hanle2,hanle3,hanle4,hanle5,hanle6,hanle7,hanle8}. Various extrinsic mechanisms have been proposed to explain this discrepancy, including lattice deformations \cite{extrinsic1}, metallic adsorbates \cite{extrinsic2,extrinsic3}, or magnetic resonances \cite{extrinsic4,extrinsic5}. In experiments and theories that assume the DP or EY mechanism, the loss of spin polarization is controlled by momentum scattering and is applicable when $\tau_s \gg \tau_p$. However, impurity scattering might cease to dominate the spin relaxation in high-mobility materials. To date, there is a lack of theoretical description of spin decoherence in this regime, where charges can propagate ballistically over long distances.

This Letter presents a study of spin dynamics in Rashba SOC materials in the absence of momentum scattering. In this regime, the spin lifetime is limited by dephasing arising from a combination of energy broadening and nonuniform spin precession. Using graphene as an example, we show that its particular band structure can yield short spin lifetimes, even for modest values of broadening and SOC. The spin dephasing is also shown to be strongly anisotropic, and the spin lifetime exhibits a characteristic dependence on the charge density, mediated by the spin-split band structure. Taken together, these features offer insight into the fundamental nature of spin dynamics in ballistic graphene, and suggest approaches to control spin lifetimes by material and device design. Beyond graphene, we briefly consider the spin dynamics of the surface state of a 3D topological insulator, and derive a temperature-dependent spin lifetime in the ballistic limit.

\textit{Spin lifetime in clean systems}. When momentum scattering is negligible, a finite spin lifetime can arise from dephasing, where an oscillating signal loses strength by mixing with other signals of different phase or frequency. In the presence of SOC, a charge carrier's spin will precess around an effective magnetic field $\vec{B}_{eff}$. If $\vec{B}_{eff}$ is energy- or momentum-dependent, and if the charge carriers are distributed in energy or momentum, the total spin signal will undergo dephasing and will decay. As a simple example, consider a system whose precession frequency varies linearly with energy, $\omega(E) = \omega_0 + \alpha E$, and whose carriers occupy a Lorentzian energy distribution, $\mathcal{L}(E) = \eta / [\pi \cdot (E^2 + \eta^2)]$, where $\eta$ is the half-width at half-maximum. The total spin signal is
\begin{equation} \label{eq:convolve}
s(t) = \mathcal{L}(E) \circ \cos(\omega(E) t) = e^{-\alpha \eta t} \cdot \cos(\omega_0 t),
\end{equation}
where $\circ$ represents the convolution integral \cite{ft1}. In general, Eq. (\ref{eq:convolve}) shows that energy broadening plus nonuniform spin precession leads to a decay of the spin due to dephasing, with a decay rate $\tau^{-1}_s = \alpha\eta$. For a continuous energy distribution the decay is irreversible; the magnitude of the signal will never recover to its original value. In reality, a finite number of charge carriers will occupy a discrete set of energies, but this also yields irreversible decay if the carriers are randomly distributed in energy. The decay is not necessarily exponential, but depends on the broadening and the variation of the precession. For example, replacing the Lorentzian with a Fermi distribution gives a decay of $\xi t / \sinh(\xi t)$, where $\xi = \pi \alpha kT$ and $kT$ is the thermal energy \citep{ft2}, while a finite bias window yields a decay of $\sinc(\alpha V_{SD}t)$, where $V_{SD}$ is the source-drain bias.

\textit{Band structure of graphene with SOC}. As discussed above, a nonuniform precession can lead to spin decay in a clean system. With that in mind, we examine the band structure of graphene in the presence of SOC. Considering a single $\pi$-orbital per carbon atom, the tight-binding Hamiltonian is
\begin{equation} \label{eq:hamiltonian_tb}
\begin{split}
\hat{H} &= -t \sum_{\langle ij \rangle} c_i^\dag c_j
        + iV_R \sum_{\langle ij \rangle}
          c_i^\dag \vec{z} \cdot (\vec{s} \times \vec{d}_{ij}) c_j \\
        &+ i \frac{2}{\sqrt{3}} V_I \sum_{\langle\langle ij \rangle\rangle}
          c_i^\dag \vec{s} \cdot (\vec{d}_{kj} \times \vec{d}_{ik}) c_j,
\end{split}
\end{equation}
where $\vec{s}$ are the spin Pauli matrices, $t$ is the nearest-neighbor hopping, $V_I$ is the intrinsic SOC, and $V_R$ is the Rashba SOC, induced by a transverse electric field or substrate \cite{tb1}. Putting Eq. (\ref{eq:hamiltonian_tb}) into the spin+pseudospin basis and taking the Fourier transform yields
\begin{equation} \label{eq:hamiltonian_matrix}
\hat{H} =
\begin{bmatrix}
\beta   &   \kappa   &   0   &   i\gamma_+ \\
\kappa^\ast   &   -\beta   &   -i\gamma_-^\ast   &   0 \\
0   &   i\gamma_-   &   -\beta   &   \kappa \\
-i\gamma_+^\ast   &   0   &   \kappa^\ast   &   \beta
\end{bmatrix},
\end{equation}
where $\kappa = -te^{ik_y \cdot 2b/3} \cdot [1 + 2e^{-ik_y b} \cos(k_x a)]$, $\gamma_\pm = V_R \cdot e^{ik_y \cdot 2b/3} \cdot [1 + 2e^{-ik_y b}\cos(k_x a \pm 2\pi/3)]$, $\beta = -V_I \cdot [2\sin(2k_x a) - 4\cos(k_y b)\sin(k_x a)]$, $k_x$ and $k_y$ are the momenta along the $x$- and $y$-axes, $a = \sqrt{3}/2 \cdot a_{cc}$, $b = 3/2 \cdot a_{cc}$, and $a_{cc}$ is the carbon-carbon distance \cite{tb2,tb3}.

\begin{figure}[t]

\includegraphics[width=\columnwidth]{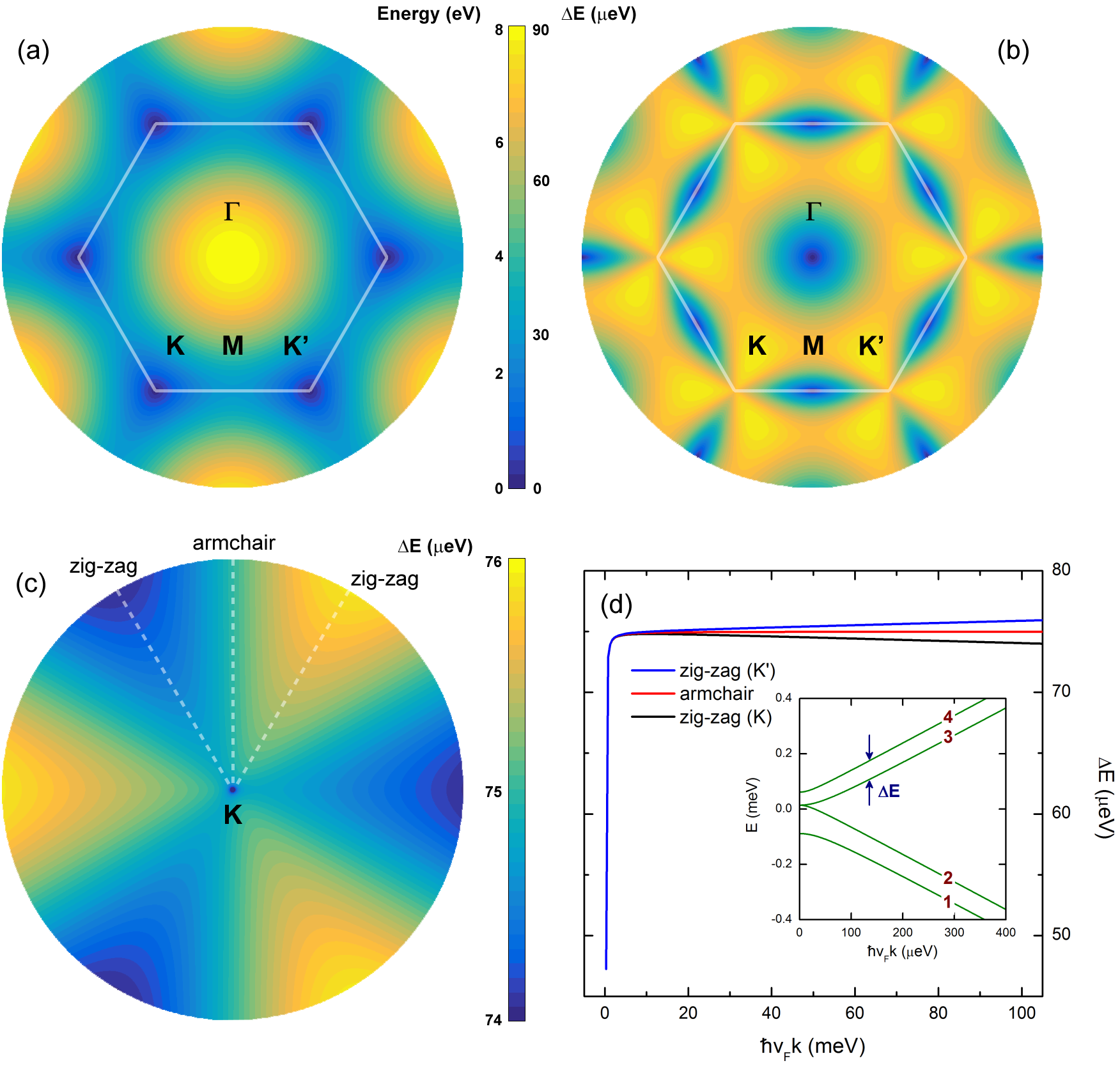}
\caption{(color online) Band structure of graphene with SOC. (a) The conduction band and (b) spin splitting of the conduction band over the entire Brillouin zone. (c) Splitting of the conduction band near the K point. (d) Splitting of the conduction band near the K and K' points for the armchair and zig-zag directions. Inset: slice of the band structure near the K point (bands are labeled 1 to 4).}
\label{fig:bandstructure}
\end{figure}

In Fig. \ref{fig:bandstructure} we plot the band structure, assuming $t$ = 2.7 eV, $V_I$ = 2.31 $\mu$eV, and $V_R$ = 25 $\mu$eV \cite{soc1, soc2, soc3, soc4, soc5}. Figure \ref{fig:bandstructure}(a) shows the conduction band over the full Brillouin zone, with Dirac cones at the corners and trigonal warping at higher energies. In Fig. \ref{fig:bandstructure}(b) we plot the spin splitting of the conduction band, which is zero at the $\Gamma$ and M points and more complex near the K and K' points. A zoom of this is shown in Fig. \ref{fig:bandstructure}(c) around the K point, indicating a nonuniform and anisotropic splitting. This is shown in more detail in Fig. \ref{fig:bandstructure}(d), where the splitting is plotted along the zig-zag and armchair directions for the K and K' valleys. Along the armchair direction, the splitting increases rapidly away from the Dirac point and saturates at a constant value. Along the zig-zag direction the splitting does not saturate, but instead varies slowly after the initial rise.

\begin{figure}[t]
\includegraphics[width=\columnwidth]{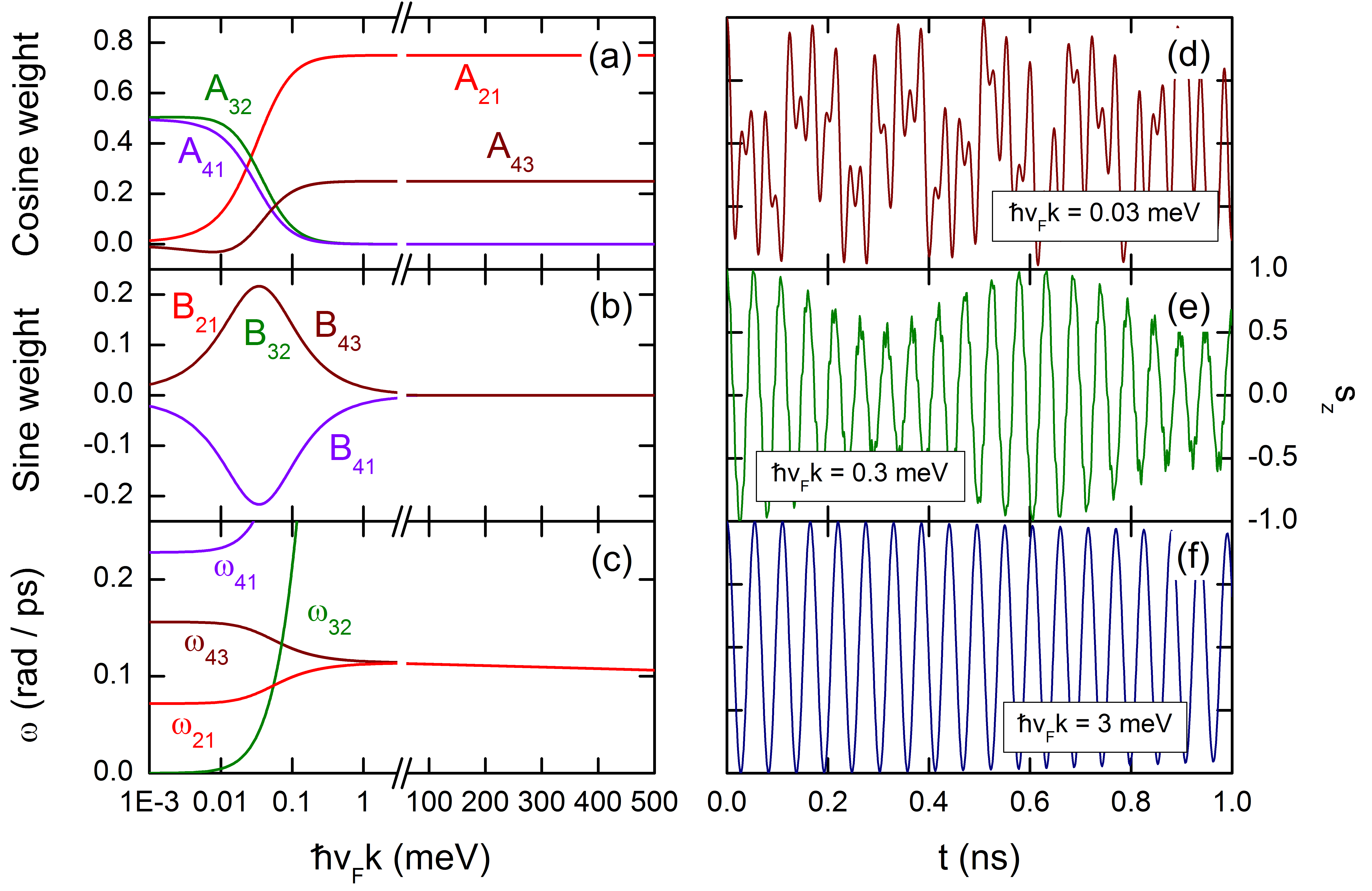}
\caption{(color online) Spin dynamics in graphene with SOC. (a) The cosine weights, (b) the sine weights, and (c) the precession frequency vs. momentum $k$, starting from the K valley and moving along the zig-zag direction. (d)-(f) Time-dependent spin polarization for selected values of $k$. The spin is projected along the $z$-axis.}
\label{fig:dynamics}
\end{figure}

\textit{Spin dynamics of graphene with SOC}. To understand the connection between the band structure and spin dephasing, we first consider the spin dynamics in a clean system. Starting with $\hat{H}|\phi_i\rangle = \epsilon_i|\phi_i\rangle$, where $\epsilon_i$ and $|\phi_i\rangle$ are the eigenvalues and eigenvectors of $\hat{H}$, the time-dependent spin polarization of an initial state $|\psi_0\rangle$ is
\begin{equation} \label{eq:spin_dynamics}
\vec{p}(t) = \sum_i \vec{A}_{ii}
           + \sum_{i>j} [\vec{A}_{ij}\cos(\omega_{ij}t)
           + \vec{B}_{ij}\sin(\omega_{ij}t)],
\end{equation}
where $\vec{A}_{ij} (\vec{B}_{ij})$ is the real (imaginary) part of $\langle\psi_0|\phi_i\rangle \langle\phi_i|\vec{s}|\phi_j\rangle \langle\phi_j|\psi_0\rangle$ and the sums run over all eigenstates at a given momentum $k$. The spin polarization consists of a constant term that depends on the polarization of each band, $\langle\phi_i|\vec{s}|\phi_i\rangle$, and an oscillating term whose frequencies are determined by the band splitting, $\omega_{ij} = (\epsilon_i - \epsilon_j) / \hbar$. The weights of the oscillating terms are determined by the spin-mediated band overlap, $\langle\phi_i|\vec{s}|\phi_j\rangle$.

As illustrated in Fig. \ref{fig:bandstructure}, $\hat{H}$ depends strongly on the momentum $k$, and therefore so will the precession. This is shown in Fig. \ref{fig:dynamics}, where we plot the weights, frequencies, and spin dynamics for $k$ along the zig-zag direction near the K point. Since each eigenstate is polarized in the $xy$-plane, we consider polarization along the $z$-axis to study the precession. In Fig. \ref{fig:dynamics}, there are two clear regimes of behavior. At large $k$, the dynamics are dominated by oscillations between the two valence bands (bands 1 and 2, see the inset of Fig. \ref{fig:bandstructure}(d)) and the two conduction bands (bands 3 and 4). In this regime, $\omega_{21}$ and $\omega_{43}$ are nearly identical, leading to the regular oscillation in Fig. \ref{fig:dynamics}(f). At smaller $k$, $\omega_{21}$ and $\omega_{43}$ diverge due to the electron-hole asymmetry induced by the intrinsic SOC, giving the beating pattern in Fig. \ref{fig:dynamics}(e). As $k \rightarrow 0$, the dominant frequencies switch from $\omega_{21}$ and $\omega_{43}$ to $\omega_{32}$ and $\omega_{41}$. Near the transition, the dynamics are governed by a combination of all frequencies, giving the complex precession in Fig. \ref{fig:dynamics}(d).

The transition between the low- and high-$k$ regimes can be understood from the eigenstates of $\hat{H}$. To illustrate we consider a continuum model, $\hat{H} = \hbar v_F \vec{\sigma} \cdot \vec{k} + \lambda_R (\vec{\sigma} \times \vec{s})$, where $v_F$ is the Fermi velocity, $\vec{\sigma}$ are the pseudospin Pauli matrices, and $\lambda_R$ is the Rashba strength. Assuming $k$ along the zig-zag ($+x$) direction, the eigenstates at large $k$ are $|\phi_j\rangle \approx [\begin{matrix} 1 & \nu_j & i \zeta_j & i \zeta_j \nu_j \end{matrix}]^T$, where $\nu_j = -1(+1)$ for bands 1 and 2 (3 and 4), and $\zeta_j = -1(+1)$ for bands 2 and 3 (1 and 4). The spin polarization of each eigenstate is $(0,\zeta_j,0)$ and the pseudospin polarization is $(\nu_j,0,0)$. From Eq. (\ref{eq:spin_dynamics}), the weights of the oscillating terms are then proportional to $\langle\phi_i|s_z|\phi_j\rangle = (1 + \nu_i \nu_j)(1 - \zeta_i \zeta_j)$. Thus, in the high-$k$ regime precession only occurs between eigenstates with the same pseudospin and opposite spin, i.e., only between the two conduction ($\omega_{43}$) or valence ($\omega_{21}$) bands. At small $k$ the Rashba term dominates and the eigenstates become $|\phi_{1,4}\rangle \approx [\begin{matrix} 0 & \mp 1 & i & 0 \end{matrix}]^T$ and $|\phi_{2,3}\rangle \approx [\begin{matrix} 1 & 0 & 0 & \pm i \end{matrix}]^T$. Here spin-pseudospin coupling is strong, as the spin-up and spin-down components of each eigenstate are located on opposite sublattices \cite{extrinsic3}. The conduction-valence bands no longer overlap, while the overlap between bands 1 and 4 (2 and 3) dominate the spin dynamics. This interband coupling, in conjunction with the broadening, is what can yield fast spin dephasing near the Dirac point.

\textit{Spin lifetime in graphene with SOC}. We can now make some predictions about dephasing-induced spin lifetimes in clean graphene. Based on the spin dynamics, dephasing should be fast near the Dirac point and slower at higher energies. We can also predict a strong anisotropy in the spin lifetime. Along the zig-zag direction, the precession away from the Dirac point varies continuously with energy, and $\tau_s$ should be constant. Along the armchair direction the precession frequency is constant, such that $\tau_s$ should diverge at high energies. For transport in all directions simultaneously, the anisotropy should produce increased dephasing due to the mixing in $k$, thus reducing $\tau_s$.

To test these predictions, we turn to numerical calculations of spin dynamics in clean graphene. We compute the time- and energy-dependent spin polarization of an initial state $|\psi_0\rangle$ as \cite{extrinsic3,eh}
\begin{equation} \label{eq:spin_numerical}
\vec{p}(E,t) = \frac{\sum_k [\langle\psi(t)|\vec{s}\delta(E-\hat{H})|\psi(t)\rangle
                             + \text{h.c.}]}
                    {2 \cdot \sum_k \langle\psi(t)|\delta(E-\hat{H})|\psi(t)\rangle},
\end{equation}
where $|\psi(t)\rangle = U(t)|\psi_0\rangle$, $U(t) = \sum_j |\phi_j\rangle \langle\phi_j| e^{-i \epsilon_j t / \hbar}$, $\delta(E-\hat{H}) = \sum_j |\phi_j\rangle \langle\phi_j| g(E-\epsilon_j)$, and $g$ is a broadening function that can be Lorentzian, a Fermi distribution, etc. The sum represents a sample over $k$-space, along a single direction or over the entire Brillouin zone, and includes both valleys. To extract $\tau_s$, we examine the time dependence of $\vec{p}(E,t)$ at each energy. In general, the complex spin dynamics near the Dirac point preclude a fit to a simple decaying cosine; instead, we define $\tau_s$ as the time when the envelope of $\vec{p}(E,t)$ falls below $e^{-1}$.

\begin{figure}[t]

\includegraphics[width=\columnwidth]{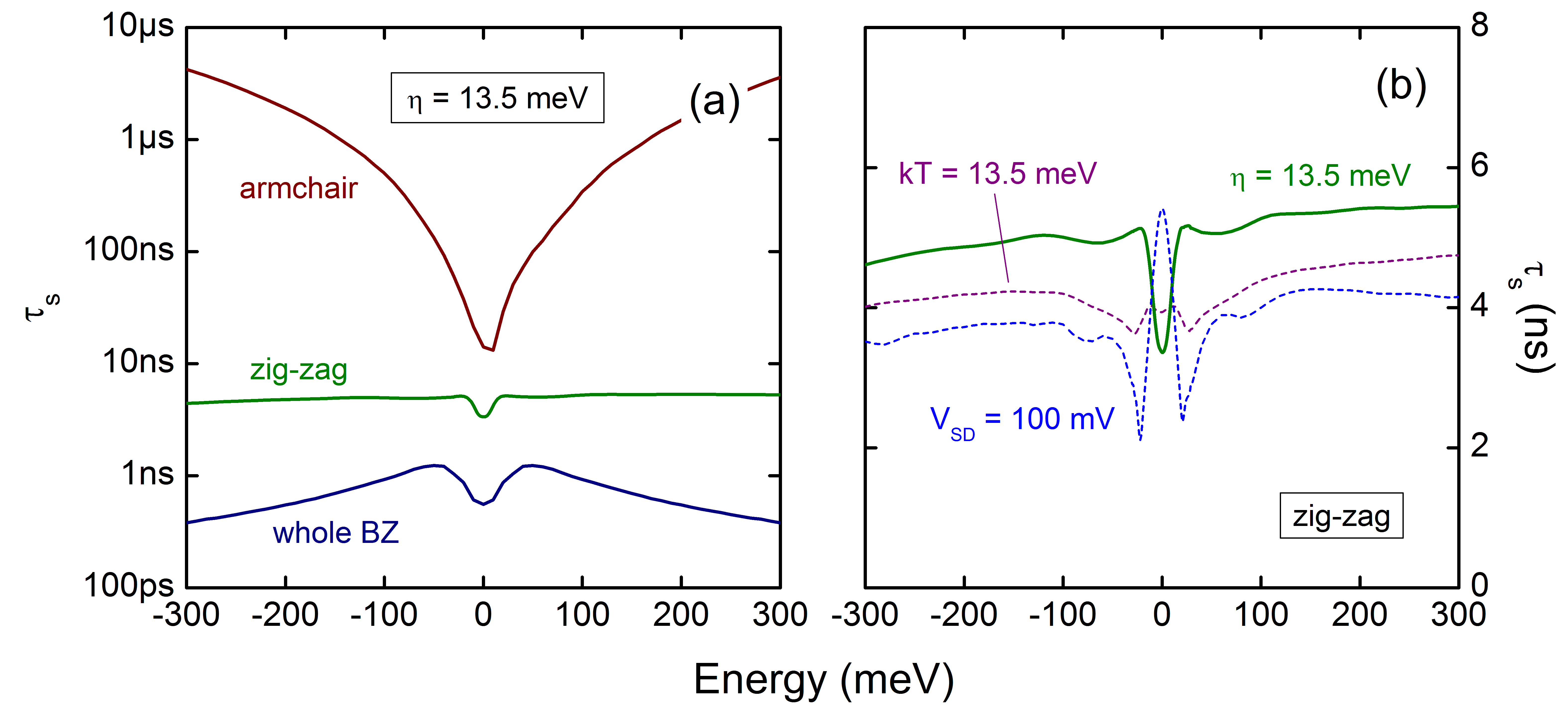}
\caption{(color online) Spin lifetime in graphene with SOC. (a) The energy-dependent spin lifetime is strongly anisotropic. (b) The spin lifetime can also be limited by thermal broadening or a finite source-drain bias (zig-zag direction).}
\label{fig:relaxation}
\end{figure}

The numerical results are shown in Fig. \ref{fig:relaxation}(a), where we plot $\tau_s$ as a function of energy, assuming Lorentzian broadening with $\eta = 13.5$ meV and polarization along the $z$-axis. Along the armchair direction, $\tau_s$ diverges with increasing energy, reaching 4 $\mu$s at 300 meV. However, near the Dirac point the dephasing limits $\tau_s$ to 14 ns. Along the zig-zag direction, $\tau_s$ saturates to 5-6 ns with a slightly lower value of 4 ns at the Dirac point. For transport in all directions, dephasing is much stronger due to the anisotropic spin dynamics, giving $\tau_s$ between 380 ps and 1.2 ns. A characteristic M-shape is observed, with the high-energy downturn of $\tau_s(E)$ resulting from the increased anisotropy of the spin splitting, as pictured in Fig. \ref{fig:bandstructure}.

The Lorentzian broadening in Fig. \ref{fig:relaxation}(a) highlights the main features of the dephasing, and the magnitude coincides with energy scales and defect densities that are common in experiments. However, this broadening is usually energy-dependent, and $\eta=13.5$ meV corresponds to an inelastic scattering time of 50 fs, much shorter than the spin precession time. Therefore, we also consider two other sources of broadening that may occur in the ballistic limit. Local and nonlocal measurements have demonstrated that the electronic temperature $T_{el}$ can be much larger than the lattice temperature, such that thermal broadening could dominate the spin dynamics without electron-phonon scattering \cite{hot1,hot2}. In two-terminal measurements a source-drain bias also serves as a source of broadening, with transport occurring over a finite energy window. In Fig. \ref{fig:relaxation}(b) we show the impact of these types of broadening for transport along the zig-zag direction. For $T_{el} = 160$ K ($kT = 13.5$ meV) or a source-drain bias of 100 mV, $\tau_s \approx 4$ ns, similar to the Lorentzian broadening. The spin dynamics are quite complex near the Dirac point, such that the energy dependence depends on the type of broadening and the specific definition of $\tau_s$. However, in general the quantitative features of Fig. \ref{fig:relaxation}(a) are independent of the type of broadening.

\textit{Discussion and conclusions}. To summarize, we have shown that the combination of energy broadening and nonuniform spin precession leads to dephasing that dictates spin lifetimes in the ballistic regime. It is important to note that without precession there will be no dephasing. As shown in Eq. (\ref{eq:spin_dynamics}), the spin signal consists of static and oscillating components, and only the oscillating components decay in time. Experimentally, spin is usually injected in the plane and perpendicular to the transport direction, resulting in an infinite lifetime in the ballistic limit. Thus, for dephasing to occur, an extrinsic effect is needed to rotate the spin polarization and allow for precession. This suggests that non-local Hanle measurements, which employ a perpendicular magnetic field \cite{hanle1}, bring a source of dephasing not present in two-terminal experiments \cite{twoterm1}. This difference may not matter for fast momentum scattering, but could become important in very clean samples. Fig. \ref{fig:relaxation}(a) showed that the spin lifetime is highly anisotropic, indicating that graphene spintronic devices could be optimized with proper lattice orientation, or by collimating the injected current \cite{Park2008}. The anisotropy is likely washed out in most experiments, with $\tau_p \approx 10$ fs much smaller than the typical spin precession time ($\sim50$ ps in this work). The results of Fig. \ref{fig:relaxation}(b) illustrate the impact of hot carriers and finite bias on spin dephasing, and suggest that these effects can impose fundamental limits on $\tau_s$. For a source-drain bias of 100 mV we find $\tau_s \approx 4$ ns along the zig-zag direction. From Eq. (\ref{eq:convolve}), $\tau_s$ scales inversely with the Rashba strength and the bias, so reducing $V_R$ to 10 $\mu$eV and $V_{SD}$ to 10 mV would yield a lifetime of 100 ns. Note that due to the ballistic transport, the spin relaxation length would be very long.

The generality of spin dephasing can also be appreciated by studying the spin dynamics of the surface state of a 3D topological insulator. In the simplest approximation, this state is characterized by a single Dirac cone with the Hamiltonian $\hat{H} = \hbar v_F (\hat{z} \times \vec{\sigma}) \cdot \vec{k}$. Considering thermal broadening and applying Eqs. (\ref{eq:convolve}) and (\ref{eq:spin_dynamics}), the out-of-plane spin dynamics are $p_z(t) = \xi t / \sinh(\xi t) \cdot \cos(\omega_0 t)$, with $\xi = 2\pi kT / \hbar$ and $\omega_0 = 2E_F / \hbar$. This yields a lifetime of $\tau_s = \upsilon / T$, with $\upsilon = 3.3$ ps-K, giving $\tau_s = 11$ fs at room temperature. Recent theoretical work on topological surface states found $\tau_s = \tau_{tr}$, where $\tau_{tr}$ is the charge transport time, suggesting that the charge transport properties can be read directly from the spin dynamics \cite{ti_1}. However, this may only be true in the low temperature limit, while thermal broadening may dominate the spin lifetime at higher temperatures.

To conclude, we have shown that even for reasonable values of broadening (meV) and Rashba SOC ($\mu$eV), spin lifetimes in clean graphene can still be very short. This suggests that spin lifetimes in Rashba SOC materials, in the absence of extrinsic effects, may have a fundamental limit related to the intrinsic bandstructure. While these results were for the limiting case of ballistic transport, they can offer insight into the nature of dephasing and spin lifetimes in disordered systems. They could also impact the observability of phenomena such as the anomalous Hall effect \cite{ahe_1}. Beyond graphene, the approach and methodology are applicable to other SOC materials, including 2DEGs \cite{2deg_1, 2deg_2, 2deg_3}, 2D transition metal dichalcogenides \cite{mos2_1, mos2_2, mos2_3}, or topological insulators \cite{ti_1, ti_2}.

\begin{acknowledgments}
The authors thank the referees for insightful comments, and are indebted to D. Soriano, J.E. Barrios-Vargas, S. Valenzuela, J. Fabian, and D. Van Tuan for useful discussions. This work is supported by the European Union Seventh Framework Programme under grant agreement 604391 Graphene Flagship, the Severo Ochoa Program (MINECO, Grant SEV-2013-0295), the Spanish Ministry of Economy and Competitiveness (MAT2012-33911), and Secretar\'{i}a de Universidades e Investigaci\'{o}n del Departamento de Econom\'{i}a y Conocimiento de la Generalidad de Catalu\~{n}a.
\end{acknowledgments}



\begin{thebibliography}{99}


\bibitem{RashbaSOC}
Y. A. Bychkov and E. I. Rasbha, Pis'ma Zh. Eksp. Teor. Fiz. {\bf 39}, 66 (1984) [JETP Lett. {\bf 39}, 78 (1984)].


\bibitem{DattaDas}
S. Datta and B. Das, Appl. Phys. Lett. {\bf 56}, 665 (1990).


\bibitem{SHE1}
J. Sinova, D. Culcer, Q. Niu, N.A. Sinitsyn, T. Jungwirth, and A.H. MacDonald, Phys. Rev. Lett. {\bf 92}, 126603 (2004).

\bibitem{SHE2}
J. Sinova, S.O. Valenzuela, J. Wunderlich, C.H. Back, and T. Jungwirth, Rev. Mod. Phys. {\bf 87}, 1213 (2015).

\bibitem{SHE3}
A. Reynoso, G. Usaj, and C.A. Balseiro, Phys. Rev. B {\bf 73}, 115342 (2006).


\bibitem{RashbaReview1}
A. Manchon, H. C. Koo, J. Nitta, S. M. Frolov, and R. A. Duine, Nat. Mater. {\bf 14}, 871 (2015).


\bibitem{theory1}
C. Ertler, S. Konschuh, M. Gmitra, and J. Fabian, Phys. Rev. B {\bf 80}, 041405 (2009).

\bibitem{theory2}
D. Huertas-Hernando, F. Guinea, and A. Brataas, Phys. Rev. Lett. {\bf 103}, 146801 (2009).

\bibitem{theory3}
Y. Zhou and M.W. Wu, Phys. Rev. B {\bf 82}, 085304 (2010).


\bibitem{mos2_1}
H. Ochoa and R. Roldan, Phys. Rev. B {\bf 87}, 245421 (2013).

\bibitem{mos2_2}
L. Wang and M.W. Wu, Phys. Rev. B {\bf 89}, 115302 (2014).

\bibitem{mos2_3}
A. Kormanyos, V. Z\'{o}lyomi, N.D. Drummond, P. Rakyta, G. Burkard, and V.I. Fal'ko, Phys. Rev. B {\bf 88}, 045416 (2013).


\bibitem{mos2_4}
D. Xiao, G.-B. Liu, W. Feng, X. Xu, W. Yao, Phys. Rev. Lett. {\bf 108}, 196802 (2012).


\bibitem{graphene_prox1}
M. Gmitra and J. Fabian, Phys. Rev. B {\bf 92}, 155403 (2015).


\bibitem{flagship}
S. Roche {\it et al.}, 2D Mater. {\bf 2}, 030202 (2015).


\bibitem{dp1}
M.I. Dyakonov and V.I. Perel, Sov. Phys. Solid State {\bf 13}, 3023 (1972).


\bibitem{ey1}
P.G. Elliot, Phys. Rev. {\bf 96}, 266 (1954).

\bibitem{ey2}
Y. Yafet, {\it Solid State Physics}, eds. F. Seitz and D. Turnbull (Academic, New York, 1963), Vol. 13.


\bibitem{ey_graphene}
H. Ochoa, A. H. Castro Neto,  F. Guinea, Phys. Rev. Lett. {\bf 108}, 206808 (2012).


\bibitem{soc1}
H. Min, J.E. Hill, N.A. Sinitsyn, B.R. Sahu, L. Kleinman, and A.H. MacDonald, Phys. Rev. B {\bf 74}, 165310 (2006).

\bibitem{soc2}
D. Huertas-Hernando, F. Guinea, and A. Brataas, Phys. Rev. B {\bf 74}, 155426 (2006).

\bibitem{soc3}
Y. Yao, F. Ye, X.-L. Qi, S.-C. Zhang, and Z. Fang, Phys. Rev. B {\bf 75}, 041401 (2007).

\bibitem{soc4}
M. Gmitra, S. Konschuh, C. Ertler, C. Ambrosch-Draxl, and J. Fabian, Phys. Rev. B {\bf 80}, 235431 (2009).

\bibitem{soc5}
S. Konschuh, M. Gmitra, and J. Fabian, Phys. Rev. B {\bf 82}, 245412 (2010).


\bibitem{hanle1}
N. Tombros, C. Jozsa, M. Popinciuc, H.T. Jonkman, and B.J. van Wees, Nature {\bf 448}, 571 (2007).

\bibitem{hanle2}
C. J\'{o}zsa, T. Maassen, M. Popinciuc, P.J. Zomer, A. Veligura, H.T. Jonkman, and B J. van Wees, Phys. Rev. B {\bf 80}, 241403 (2009).

\bibitem{hanle3}
A. Avsar {\it et al.}, Nano Lett. {\bf 11}, 2363 (2011).

\bibitem{hanle4}
W. Han and R.K. Kawakami, Phys. Rev. Lett. {\bf 107}, 047207 (2011).

\bibitem{hanle5}
T. Maassen, J.J. van den Berg, N. IJbema, F. Fromm, T. Seyller, R. Yakimova, and B. J. van Wees, Nano Lett. {\bf 12}, 1498 (2012).

\bibitem{hanle6}
I. Neumann, J. Van de Vondel, G. Bridoux, M.V. Costache, F. Alzina, C.M. Sotomayor Torres, and S.O. Valenzuela, Small {\bf 9}, 156 (2013).

\bibitem{hanle7}
M.H.D. Guimar\~{a}es, P.J. Zomer, J. Ingla-Ayn\'{e}s, J.C. Brant, N. Tombros, and B.J. van Wees, Phys. Rev. Lett. {\bf 113}, 086602 (2014). 

\bibitem{hanle8}
M.V. Kamalakar, C. Groenveld, A. Dankert, and S.P. Dash, Nat. Commun. {\bf 6}, 6766 (2015).


\bibitem{extrinsic1}
A.H. Castro Neto and F. Guinea, Phys. Rev. Lett. {\bf 103}, 026804 (2009).

\bibitem{extrinsic2}
C. Weeks, J. Hu, J. Alicea, M. Franz, and R. Wu, Phys. Rev. X {\bf 1}, 021001 (2011).

\bibitem{extrinsic3}
D.V. Tuan, F. Ortmann, D. Soriano, S.O. Valenzuela, and S. Roche, Nat. Phys. {\bf 10}, 857 (2014).

\bibitem{extrinsic4}
D. Kochan, M. Gmitra, and J. Fabian, Phys. Rev. Lett. {\bf 112}, 116602 (2014).

\bibitem{extrinsic5}
D. Kochan, S. Irmer, M. Gmitra, and J. Fabian, Phys. Rev. Lett. {\bf 115}, 196601 (2015).


\bibitem{ft1}
T. Butz, {\it Fourier Transformation for Pedestrians} (Springer International, Cham, Switzerland, 2006).

\bibitem{ft2} 
G. Bevilacqua, arXiv:1303.6206.


\bibitem{tb1}
C.L. Kane and E.J. Mele, Phys. Rev. Lett. {\bf 95}, 226801 (2005).

\bibitem{tb2}
M. Zarea and N. Sandler, Phys. Rev. B {\bf 79}, 165442 (2009).

\bibitem{tb3}
R. van Gelderen and C.M. Smith, Phys. Rev. B {\bf 81}, 125435 (2010).


\bibitem{eh}
D. V. Tuan, F. Ortmann, A.W. Cummings, D. Soriano, and S. Roche, Sci. Rep. {\bf 6}, 21046 (2016).


\bibitem{hot1}
A.C. Betz, S.H. Jhang, E. Pallecchi, R. Ferreira, G. F\`{e}ve, J-M. Berroir, and B. Pla\c{c}ais, Nat. Phys, {\bf 9}, 109 (2013).

\bibitem{hot2}
J.F. Sierra, I. Neumann, M.V. Costache, and S.O. Valenzuela, Nano Lett. {\bf 15}, 4000 (2015).


\bibitem{twoterm1}
B. Dlubak {\it et al.}, Nat. Phys. {\bf 8}, 557 (2012).


\bibitem{Park2008}
C.-H. Park, Y.-W. Son, L. Yang, M.L. Cohen, and S.G. Louie, Nano Lett. {\bf 8}, 2920-2924 (2008).


\bibitem{ahe_1}
Z. Wang, C. Tang, R. Sachs, Y. Barlas, and J. Shi, Phys. Rev. Lett. {\bf 114}, 016603 (2015).


\bibitem{2deg_1}
S. LaShell, B.A. McDougall, and E. Jensen, Phys. Rev. Lett. {\bf 77}, 3419 (1996).

\bibitem{2deg_2}
Yu.M. Koroteev, G. Bihlmayer, J.E. Gayone, E.V. Chulkov, S. Bl\"{u}gel, P.M. Echenique, and Ph. Hofmann, Phys. Rev. Lett. {\bf 93}, 046403 (2004).

\bibitem{2deg_3}
C. Zhao, J. Li, Y. Yu, H. Ni, Z. Niu, and X. Zhang, Appl. Phys. Lett. {\bf 104}, 052411 (2014).


\bibitem{ti_1}
X. Liu and J. Sinova, Phys. Rev. Lett. {\bf 111}, 166801 (2013).

\bibitem{ti_2}
E. Wang, P. Tang, G. Wan, A.V. Fedorov, I. Miotkowski, Y.P. Chen, W. Duan, and S. Zhou, Nano Lett. {\bf 15}, 2031 (2015).


\end{thebibliography}
\end{document}